\documentclass[12pt]{article}
    \newcommand{\be}[1]{\begin{equation}\label{#1}}
    \newcommand{\ba}[1]{\begin{eqnarray}\label{#1}}
    \newcommand{\ep}[1]{\epsilon_{#1}}

    \newcommand{\pa}[1]{\left(#1\right)}
    \newcommand{\paq}[1]{\left[#1\right]}

    \newcommand{\M}{{\rm M_{\rm P}}}

    \def\ee{\end{equation}}
    \def\ea{\end{eqnarray}}

\usepackage{graphicx,latexsym}
\usepackage{graphicx,epsf}
\usepackage{color}
\usepackage{epsfig}
\usepackage{amsmath}
\usepackage[T1]{fontenc}
\usepackage[utf8]{inputenc}
\usepackage{authblk}
\begin{document}
\title{Quantum Gravity and the Large Scale Anomaly}

\author[1,2]{Alexander Y. Kamenshchik\thanks{Alexander.Kamenshchik@bo.infn.it}}
\author[1]{Alessandro Tronconi\thanks{Alessandro.Tronconi@bo.infn.it}}
\author[1]{Giovanni Venturi\thanks{Giovanni.Venturi@bo.infn.it}}
\affil[1]{Dipartimento di Fisica e Astronomia and INFN, Via Irnerio 46,40126 Bologna,
Italy}
\affil[2]{L.D. Landau Institute for Theoretical Physics of the Russian
Academy of Sciences, Kosygin str. 2, 119334 Moscow, Russia}
\date{}
\renewcommand\Authands{ and }
\maketitle

\begin{abstract}
The spectrum of primordial perturbations obtained by calculating the quantum gravitational corrections to the dynamics of scalar perturbations is compared with Planck 2013 and BICEP2/{\it Keck Array} public data. The quantum gravitational effects are calculated in the context of a Wheeler-De Witt approach and have quite distinctive features. We constrain the free parameters of the theory by comparison with observations. 
\end{abstract}

\section{Introduction}
The spectra of the inhomogeneities in the cosmic microwave background (CMB) contain much information on the evolution of the Universe from its early stages until now. The latest Planck mission \cite{planck} measured such inhomogeneities and their statistical properties with great accuracy thus improving the present constraints on cosmological evolution. Planck results indicate that the so called Standard Cosmological model can fit CMB data well on assuming that the spectra of the primordial cosmological perturbations at the onset of the hot big bang are given by $\mathcal P\sim A (k/k_*)^{m}$ where $k_*$ is a suitable pivot scale. Such a power law dependence and a red tilted spectrum in the scalar sector may be produced by an inflationary era \cite{inflation} driven by a single, minimally coupled scalar field with a flat potential \cite{Stewart:1993bc}.\\ 
Planck data confirm the mild evidence, already hinted at by WMAP \cite{WMAP} and other independent datasets, of a power suppression of the temperature anisotropies spectrum on large scales. Despite that this evidence may well be explained as an effect of cosmic variance, it is still worth investigating the intriguing possibility that it is associated with some physical effect.\\
Various attempts have been made in order to physically justify the power suppression on large scales \cite{powerloss}. In particular, it was observed that quantum gravitational effects may affect the large scales of the primordial spectra of perturbations \cite{quantumloss}. Such an effect is naively explained by the fact that the largest scales we observe in the CMB are those which first exit the horizon during inflation (and re-enter at later times) and are thus more affected by large energy/curvature effects than the shortest scales.\\
Similar models for the evolution of primordial cosmological perturbations in the framework of quantum cosmology have been investigated recently \cite{quantumloss}. In particular in \cite{Kamenshchik:2014kpa} the quantum gravitational corrections to the single, minimally coupled scalar field inflation are obtained through a Born-Oppenheimer decomposition \cite{BO} and without specifying any further assumption on the inflationary potential except for the requirement that it determines the slow-roll (SR) evolution. These corrections affect the infrared part of the spectra and lead to an amplification or a suppression depending on the vacuum choice, the value of the SR parameters and the number of e-folds of inflation. More interestingly they depend on the wavenumber $k$ and scale as $k^{-3}$ in both the scalar and the tensor sectors.\\
In this paper the predictions of this model are compared with observations and the free parameters of model are constrained. The paper is organized as follows: in Section 2 we review the basic formalism introduced in \cite{Kamenshchik:2014kpa}; in Section 3 we present the modified primordial spectra obtained on calculating the quantum gravitational corrections; in Section 4 we discuss our analysis, we specify the dataset compared with the theoretical predictions and the free parameters involved in data fitting, we report and comment the Monte Carlo results; finally the conclusions are drawn in Section 5.


\section{Formalism}
The quantum gravitational effects on the spectra of the primordial perturbations $\hat v_k$ \cite{MukMald} have been obtained in a Wheeler-De Witt approach to quantum gravity \cite{DeWitt}. These effects can be evaluated to the first order in $\M^{-2}$ on solving the following differential equation 
\be{qfx}
\frac{d^{3}p}{d\eta^{3}}+4\omega^{2}\frac{dp}{d\eta}+2\frac{d\omega^{2}}{d\eta}p+\Delta_p=0
\ee
for the two-point function 
\be{two}
p(\eta)\equiv\langle 0|\hat v_k^{2}|0\rangle
\ee
where the quantum gravitational corrections are given by
\ba{qfxP}
\!\!\!\!\!\Delta_p&\!\!\!\!\!=\!\!\!\!\!&-\frac{1}{\M^{2}}\frac{d^{3}}{d\eta^{3}}\frac{\left(p'^{2}+4\omega^{2}p^{2}-1\right)}{4 a'^{2}}+\frac{1}{\M^{2}}\frac{d^{2}}{d\eta^{2}}\frac{p'\left(p'^{2}+4\omega^{2}p^{2}+1\right)}{4 pa'^{2}}\nonumber\\
&&+\frac{1}{\M^{2}}\frac{d}{d\eta}\left\{\frac{1}{8a'^{2}p^{2}}\left[\left(1-4\omega^{2}p^{2}\right)^{2}+2p'^{2}\left(1+4\omega^{2}p^{2}\right)+p'^{4}\right]\right\}\nonumber\\
&&-\frac{1}{\M^{2}}\frac{\omega\omega'\left(p'^{2}+4\omega^{2}p^{2}-1\right)}{a'^{2}}.
\ea
$\eta$ is the conformal time, $a$ is the scale factor and a prime means a derivative w.r.t. $\eta$.
In the scalar sector $\hat v_k$ is the canonically quantized Mukhanov variable, $\omega^2=k^2-\frac{z''}{z}$ with $z=\phi_0'/H$, $\phi_0$ is the homogeneous inflaton field and $H$ is the Hubble parameter. In the tensor sector $\hat v_k$ is the canonically quantized rescaled tensor perturbation and $\omega^2=k^2-\frac{a''}{a}$. Let us note that the above master equation (\ref{qfx}) is only valid within a perturbative approach and $\Delta_p$ is the expression for the quantum gravitational corrections up to order $\M^{-2}$ to the evolution of the Bunch-Davies (BD) vacuum \cite{BD}. The quantum gravitational correction term $\Delta_p$ is associated with ``non-adiabatic'' transitions between eigenstates of the invariant vacuum (BD) and higher eigenstates of the invariants \cite{FVV}.\\
For realistic inflationary models, with a slow rolling inflaton field, the equation (\ref{qfx}) can be further simplified to the first order in the SR approximation and one has
\be{omegas}
\omega^2=k^{2}-\frac{2\pa{1+3\epsilon_{SR}-\frac{3}{2}\eta_{SR}}}{\eta^{2}}
\ee
for the scalar perturbations and
\be{omegat}
\omega^2=k^{2}-\frac{2\pa{1+\frac{3}{2}\epsilon_{SR}}}{\eta^{2}}
\ee
for the tensor perturbations with 
\be{SRlit}
\epsilon_{SR}\equiv -\frac{\dot H}{H^2}\;\;{\rm and}\;\; \eta_{SR}\equiv-\frac{\ddot \phi_0}{H\dot \phi_0}
\ee
where a dot means a derivative w.r.t. the proper time $dt\equiv a d\eta$.
Given the expressions (\ref{omegas}), (\ref{omegat}) we observe that the results for the scalar sector can be transformed into the results for the tensor sector by simply replacing $\eta_{SR}$ with $\epsilon_{SR}$.


\section{Quantum gravitational corrections}
The effect of $\Delta_p$ on the evolution of the two-point function $p$ is that of adding to the standard BD solution $p_{BD}$ (leading to the standard power-law expression for the primordial spectra) a contribution of order $\M^{-2}$. Furthermore, to first order in the quantum gravitational corrections, the perturbative approach adopted still leaves the freedom of considering a slightly modified BD prescription providing the standard vacuum is recovered when $\M\rightarrow \infty$. 
In the scalar sector the final expression for $p$ has the following form in the long wavelength limit:
\be{papprox}
p^{(L)}\simeq p_{0}^{(L)}\paq{1+\frac{H^2}{\M^2}\frac{\bar k^3}{k^3}\pa{\frac{\delta_k}{k}+\frac{17}{9}-\frac{7}{18}\pa{-k\eta}^{-2\pa{\ep{SR}-\eta_{SR}}}}}.
\ee
where $\bar k$ is an unspecified reference wave number. The appearance of $\bar k$ in the quantum corrections can be traced back to the three volume integral in the original action for the homogeneous inflaton-gravity system plus perturbations \cite{Kamenshchik:2014kpa}. Such a volume, on a spatially flat homogeneous space-time, is formally infinite and consequently the value of $\bar k$ remains undetermined. Naively one may argue that $\bar k$ is related to an infrared problem (divergence) and indeed, in the literature, its value is taken to be the infrared cut-off for the perturbations, namely the largest observable scale in the CMB. The parameter $\delta_k$ in (\ref{papprox}) is related to the initial conditions for the quantum state describing the evolution of the perturbations. Such initial conditions are fixed by the BD prescription at the unperturbed level but in principle a departure from pure BD of the order $\M^{-2}$ is consistent with the approach. In general $\delta_k$ may depend on $k$ and, in particular, on setting the SR parameters to zero one recovers the exact BD solution in de Sitter for $\delta_k=-k/2$.\\
The time dependent contribution $-k\eta$ in (\ref{papprox}) is evaluated at the end of inflation. The wavenumber $k$ necessarily refers to the scales, around the pivot scale $k_*$, which are probed by CMB and exited from the horizon $N_*\sim 60$ e-folds before inflation ends. Its contribution to (\ref{papprox}) is
\ba{keta}
\pa{-k\eta}^{-2\pa{\ep{SR}-\eta_{SR}}}&\simeq&\pa{\frac{k}{a_{\rm end} H}}^{-2\pa{\ep{SR}-\eta_{SR}}}=\pa{\frac{k}{k_*}\frac{k_*}{a_*{\rm e}^{N*}H}}^{-2\pa{\ep{SR}-\eta_{SR}}}\nonumber\\
&\simeq&{\rm e}^{2N_*\pa{\ep{SR}-\eta_{SR}}}
\ea 
and may well lead to a contribution of $\mathcal{O}\pa{1}$ for reasonable values of the SR parameters of the order of 1 per cent. Let us note that the last equality in (\ref{papprox}) is strictly valid for the modes very close to the pivot scale $k\sim k_*=a_* H$. Away from the pivot scale small deviations proportional to the SR parameters, $-2(\ep{SR}-\eta_{SR})\ln \pa{\frac{k}{k_*}}$, are neglected. 
Depending on the SR parameters, on $N_*$ and on the vacuum choice ($\delta_k$) the quantum corrections may lead to a power loss or a power increase for large scales which can be generically parametrized in the following form:
\be{pparam}
p^{(L)}\simeq p_{0}^{(L)}\paq{1\pm q\pa{\frac{k_*}{k}}^3}
\ee
where the expression (\ref{pparam}) is simply obtained on comparing with (\ref{papprox}).
An analogous parametrization holds for the tensor sector with a different $q$.\\
\subsection{Extrapolation beyond NLO}
The parametrization of the primordial spectra by (\ref{pparam}) is still not suitable for comparison with observations. In the $k\ll k_*$ limit the quantum gravitational corrections to $p_0$ are either negative or very large (infinite in the $k\rightarrow 0$ limit). Such an apparently pathological behavior is simply the consequence of the perturbative technique employed to evaluate the corrections. 
One may hope that resummation to all orders leads to a finite result.
In any case we are not allowed to extend the validity of the perturbative corrections up to $\mathcal{O}\pa{1}$.\\
Thus, instead of introducing a sharp cut-off on the NLO expressions for the modified spectra by multiplying $q$ by an ad hoc step function which keeps the correction small but leads to a discontinuous spectrum, we interpolate our expression through a well defined function with a finite and reasonable behavior in the $k\rightarrow 0$ limit. Such a function, which must reproduce (\ref{pparam}) when $q\pa{k_*/k}^3\ll 1$, may be regarded as a resummation of the perturbative series.\\
In order to restrict the number of parameters which will be fitted by the comparison with the data and still allow for different limits when $k\rightarrow 0$ we consider the following parametrization:
\be{extparam}
p^{(L)}\simeq p_{0}^{(L)}\frac{1+\tilde q_1\pa{\frac{k_*}{k}}^3}{1+\tilde q_2\pa{\frac{k_*}{k}}^3}\sim p_{0}^{(L)}\paq{1+\pa{\tilde q_1-\tilde q_2}\pa{\frac{k_*}{k}}^3}.
\ee
where one more parameter w.r.t. (\ref{pparam}) has been added in order to obtain a regular expression for $k$ small. 
Let us note that the above modifications are substantially different from considering a running spectral index $\alpha_s$ such as
\be{running}
p^{(L)}\simeq p_{0}^{(L)}\pa{\frac{k_*}{k}}^{-\frac{\alpha_s}{2}\ln\pa{\frac{k}{k_*}}}
\ee
Indeed in the latter case the standard power law dependence is affected at both large and small scales and, in particular, a negative running would lead to a zero amplitude in the $k\rightarrow 0$ limit and a smaller amplitude w.r.t. simple power law when $k\gg k_*$. The modified spectrum (\ref{extparam}) reduces to the power law case when $k\gg k_*$ and may lead to a non zero amplitude when $k\rightarrow 0$ depending on the choice of the parameters $\tilde q_{1,2}$. 
\begin{table}[t]
\caption{Varying parameters}
\vspace{0.1 cm}
\centering
\begin{tabular}{c | c | c | c |  c | c }
\hline\hline
$\ln\pa{10^{10}A_s}$ & $n_s$ & $r$ & $\alpha_s$ & $q_1$ & $q_2$  \\ 
[1ex]
\hline
$[2.7,4.0]$		& $[0.9,1.1]$ 		& [0,0.8] 		& [-0.1,0.1] 		& [0,1] 		& [0,0.5]			\\ 
[1ex]
\hline
\end{tabular}
\label{tab1}
\end{table}

\section{Data Analysis}
In this section we report the comparison between the theoretical predictions given by (\ref{extparam}) and Planck 2013 \cite{planck} and BICEP2/{\it Keck Array} \cite{bicep} datasets. The analysis is performed using the Markov Chain Monte Carlo (MCMC) code {\it COSMOMC} \cite{cosmomc} which has been properly modified to take into account the estimated quantum gravitational effects.\\
Let us note that BD vacuum in the tensor sector (see \cite{quantumloss} for more details) gives a power increase at large scales in the tensor spectrum. Such an increase would be counterbalanced by a loss of power in the scalar sector as far as temperature correlations are concerned. One may parametrize such a power increase in a suitable way just as we did for the scalar sector in order to eliminate the divergence for small $k$ and fit the corresponding parameter with the data at our disposal. Since our main source of data comes from temperature correlations, which do not discriminate between scalar and tensor fluctuations, we neglect a priori quantum gravitational corrections in the tensor spectrum. Such a choice is a simplifying assumption done in order not to have to disentangle possible degenerate parameters. Let us note, however, that such a choice can be realized physically either by an appropriate vacuum choice, differing from a pure BD, or by a very long cutoff scale associated with tensor dynamics. Thus we limit our analysis to a subset of the more general case for which the quantum gravitational corrections affect the tensor sector in a non negligible way. The tensor spectrum is then given by the unperturbed power law expression
\be{tensor}
p_t=A_t\pa{\frac{k}{k_*}}^{n_t}.
\ee
and we assume that the LO spectra are generated by the conventional SR mechanism and single field inflation. As a consequence the consistency condition relating scalar and tensor spectral indices and the tensor to scalar ration is valid. Indeed throughout the analysis we assume that the consistency relation (already implemented in  {\it COSMOMC}) between the spectral indices and the tensor to scalar ratio:
\be{consistency}
n_t=-\frac{r}{8}\pa{2-n_s-\frac{r}{8}}
\ee
holds to the second order in the SR approximation and consequently the amplitude of the spectrum of tensor perturbations is given by $A_t=r A_s$.
We then consider a primordial scalar spectrum $p^{(L)}$ parametrized by 
\be{scparam}
p_s\simeq p_{0}^{(L)}\frac{1+\pa{1-2q_2}\pa{\frac{q_1k_*}{k}}^3}{1+\pa{\frac{q_1 k_*}{k}}^3}
\ee
where $1-2q_2$ simply fixes the limit of $p_s$ when $k\rightarrow 0$. Let us note that $q_2=0$ or $q_1=0$ correspond to the standard power-law case with no loss of power ($p_s=p_{0}^{(L)}$) and $q_2=0.5$ corresponds to zero power at $k=0$. The expression (\ref{scparam}) is a parametrization equivalent to (\ref{extparam}) with $\tilde q_1=q_1^3\pa{1-2q_2}$ and $\tilde q_2=q_1^3$ which we have found to be more convenient to be used in {\it COSMOMC}.\\
\begin{table}[t]
\caption{Models List}
\vspace{0.1 cm}
\centering
\begin{tabular}{c | l | |l|l}
\hline\hline
Model \# & Primordial spectra & Datasets & Parameters  \\ 
[1ex]
\hline
1				& Power law		& PL 	&$A_s$, $n_s$	\\ 
2				& no tensors			 	& PL+BK 	&\\
\hline
3	 			& Power law 		 	& PL &$A_s$, $n_s$, $r$		\\
4		 		& and tensors	 	& PL+BK 	&	\\
\hline
5			 	& Running spectral index 	 	& PL	&$A_s$, $n_s$, $r$, $\alpha_s$	\\
6			 	& and tensors	 	& PL+BK			&	\\
\hline
7			 	& Quantum gravitational 	 	&PL 		&$A_s$, $n_s$, $r$, $q_1$, $q_2$	\\
8				&  corrections and tensors		&PL+BK	& \\ 
[1ex]
\hline
\end{tabular}
\label{tab2}
\end{table}
Our analysis is based on Planck+WP (PL) datasets released in 2013, which include the Planck TT data and the WMAP9 polarization data, and the BICEP2/{\it Keck Array} dataset (BK) released in 2015. In particular we use {\it CAMspec}, {\it lowLike}, {\it lowl} and {\it lensing} publicly available Planck likelihoods. We find the best fit for our model with and without BK results and compare it with standard power law predictions and with predictions assuming a running spectral index (\ref{running}). In the standard power law case we either fix $r=0$ or let it vary.\\
For simplicity we obtained the best-fits for the parameters of the primordial spectra shown in Table \ref{tab1} and the parameters are taken to vary with uniform priors in the intervals indicated in the same table. The remaining parameters are fixed to the Planck best-fit in {\it base\_planck\_lowl\_lowLike\_highL} run. Let us note that the pivot scale $k_*$ is $0.05\;{\rm Mpc}^{-1}$ and is the same for both the scalar and the tensor sector. The priors for $A_s$, $n_s$, $r$ and $\alpha_s$ are those used by Planck 2013 analysis. The additional parameters $q_1$ and $q_2$ are chosen to vary in the largest possible interval reproducing a power loss for large scales (compared to the pivot scale) with the parametrization chosen. At present our theoretical predictions are not able to constraint the value of such parameters or estimate possible allowed intervals where to let them vary (see \cite{hert} for an attempt to estimate priors from quantum gravity), thus the choice of broad enough priors seems reasonable.
In particular the prior for $q_1$ is chosen in order to search for effects at scales larger that the pivot scale $k_{*}^{-1}$, which seems reasonable given the present data, with the limiting value $q_1=0$ which reproduces the standard power-law $p_s=p_0^{(L)}$ parametrization. The prior for $q_2$ is chosen to let the parameter vary between $q_2=0$ and $q_2=1/2$ where the standard power-law is obtained independently of $q_1$. The values of $q_2$ with $q_2>1/2$, leading to an increase of power, and with $q_2<0$, leading to a physically unacceptable negative spectrum, are excluded from the analysis.\\
The different combinations of primordial spectra and datasets considered are listed in Table \ref{tab2} and the corresponding fits are labelled with an index specifying the model number in table \ref{tab2}. The best fits for the parameters we vary are presented in Table \ref{tab3} and the corresponding effective $\chi^2$ defined as $-2\ln\mathcal{L}$, where $\mathcal{L}$ is the likelihood, are listed in Tables \ref{tab4} and \ref{tab4b}. In particular in the latter tables we present the effective $\chi^2$ derived from the full dataset used in each analysis ($\chi^2_{Tot}$) and those derived from the comparison with only the {\it Commander} dataset or the BK dataset. Let us note that {\it Commander} data only regard temperature correlation (and not polarization) and are restricted to the lowest multipoles $2\le l \le 50$. Finally the differences between the total $\chi^2$ for the different cases are reported on using our model as reference. In particular the cases 1, 3 and 5 are compared with 7 and the cases 2, 4, 6 are compared with 8.
\begin{table}[t]
\caption{Monte Carlo Best-fits}
\vspace{0.1 cm}
\centering
\begin{tabular}{c | c c c c | c c }
\hline\hline
\# & $\ln\pa{10^{10}A_s}$ & $n_s$ & $r$ & $\alpha_s$ & $q_1$ & $q_2$  \\ 
[1ex]
\hline
1 			& 3.097 	& 0.957 		& - 						& - 					& -					& -						\\ 
2			& 3.097 	& 0.957 		& - 						& - 					& -					& -						\\
3			& 3.097 	& 0.958 		& $0.61 \cdot10^{-3}$ 		& - 					& -					& -						\\
4	 		& 3.097 	& 0.958 		& $0.24\cdot 10^{-1}$ 		& - 					& -					& -						\\
5		 	& 3.100 	& 0.957 		& $0.26 \cdot 10^{-1}$ 		& $-0.12\cdot 10^{-1}$ 	& -					& -					   	\\
6		 	& 3.100 	& 0.957		& $0.39 \cdot 10^{-1}$ 		& $-0.13\cdot 10^{-1}$	& -					& -						\\
7			& 3.099 	& 0.954 		& 0.17 					& - 					& $0.73\cdot 10^{-1}$	& 0.13					\\
8			& 3.098 	& 0.955 		& $0.44\cdot 10^{-1}$ 		& -	  				& $0.59\cdot 10^{-1}$	& $0.11$				 	\\ 
[1ex]
\hline
\end{tabular}
\label{tab3}
\end{table}
\begin{table}[ht]
\caption{Monte Carlo Comparison (PL)}
\vspace{0.1 cm}
\centering
\begin{tabular}{c | c c | c }
\hline\hline
\# & $\chi^2_{Commander}$ & $\chi^2_{Tot}$ & $\Delta \chi^2\equiv \chi^2_\#-\chi^2_7$  \\
[1ex]
\hline
1 			&-5.56			&  9864.2	 	& 5.9	\\ 
3			& -5.61 			&  9864.2		& 5.9     	\\
5		 	& -8.72 			&  9860.6   	& 2.3	\\
\hline
7			& -10.6 			&  9858.3	 	& 0		\\
\hline
\end{tabular}
\label{tab4}
\end{table}
\begin{table}[ht]
\caption{Monte Carlo Comparison (PL+BK)}
\vspace{0.1 cm}
\centering
\begin{tabular}{c | c c c | c }
\hline\hline
\# & $\chi^2_{Commander}$ & $\chi^2_{BK}$& $\chi^2_{Tot}$ & $\Delta \chi^2\equiv \chi^2_\#-\chi^2_8$ \\
[1ex]
\hline
2			& -5.56 	& 46.9 		&  9911.3		& 7.2	\\
4	 		& -4.66 	&  45.7		&  9910.9		& 6.8	\\
6		 	& -8.62 	& 45.3		&  9906.7 		& 2.6	\\
\hline
8			& -10.6 	& 45.3 		&  9904.1 		& 0		\\ 
\hline
\end{tabular}
\label{tab4b}
\end{table}

\subsection{Results}
The MCMC results (see Tables \ref{tab4} and \ref{tab4b}) show that the quantum gravitational modification of the standard power law form for the primordial scalar spectrum improves the fit to the data. Such improvements are much more significant w.r.t the standard modifications of the primordial spectra obtained by considering a running spectral index. 
\begin{figure}[t!]
\centering
\epsfig{file=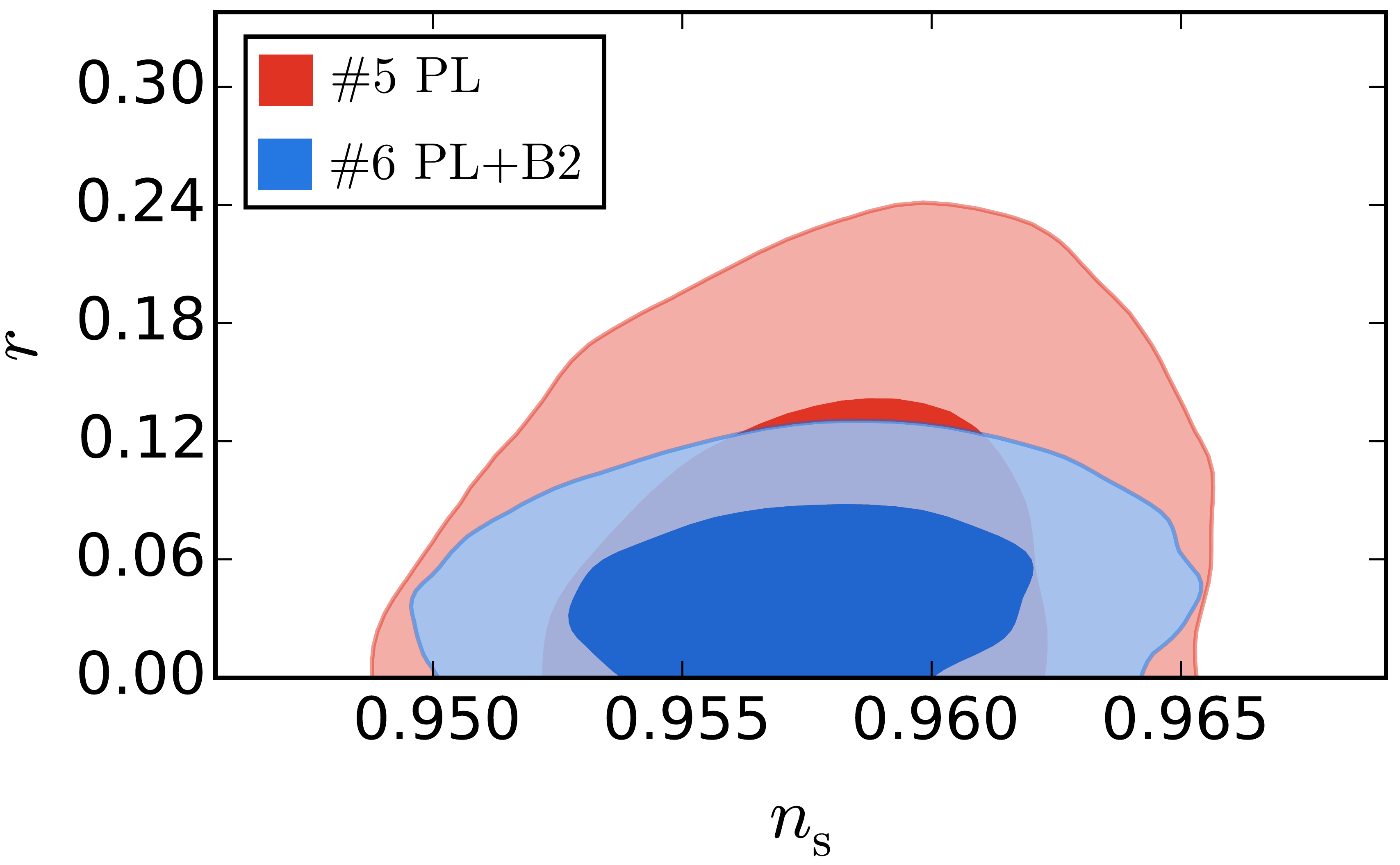, width=6.5 cm}
\caption{\it $68\%$ and $95\%$ confidence level constraints on $r$ and $n_s$ in cases $5-6$.}
\label{fig1}
\end{figure}
\begin{figure}[t!]
\centering
\epsfig{file=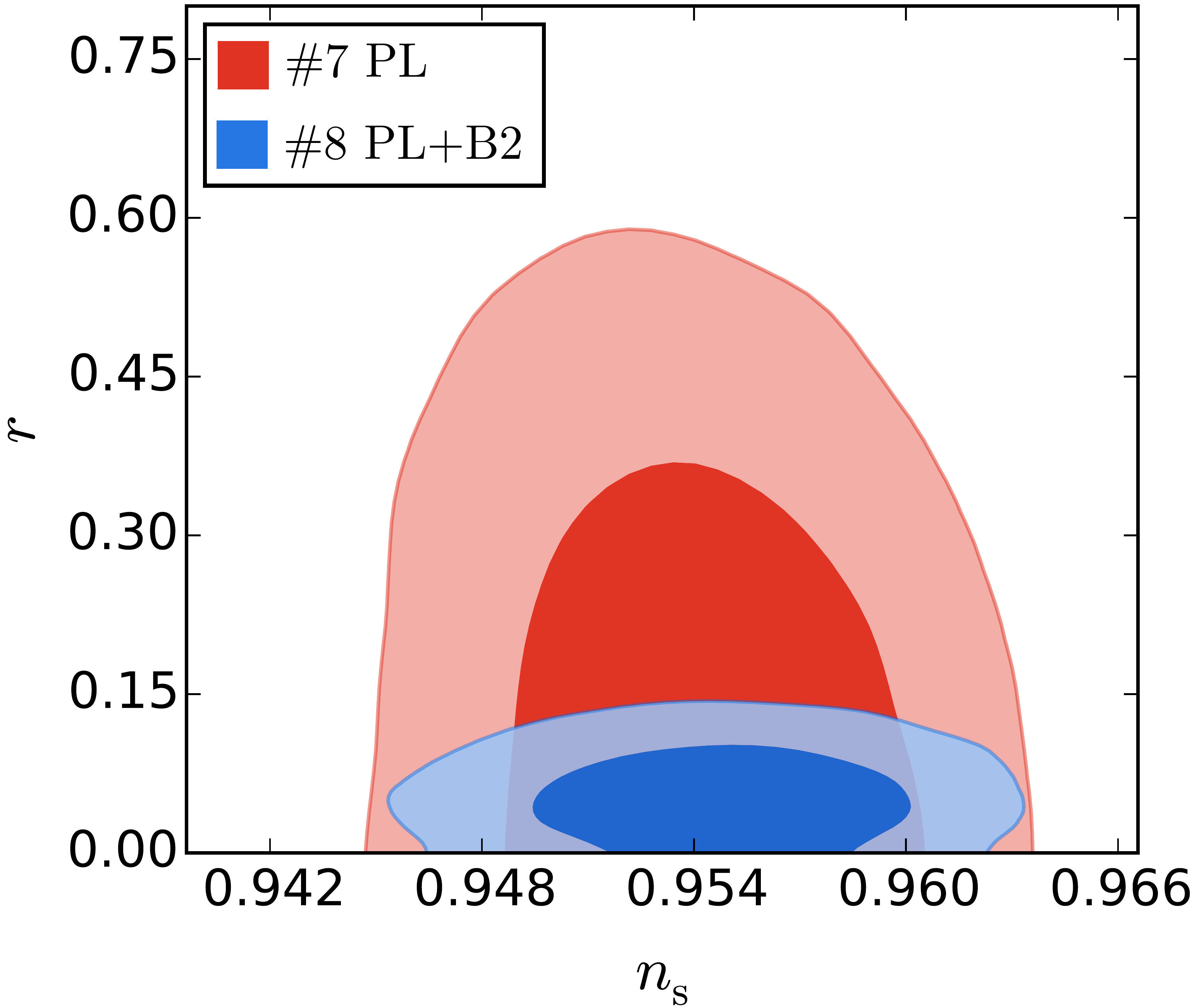, width=6.5 cm}
\epsfig{file=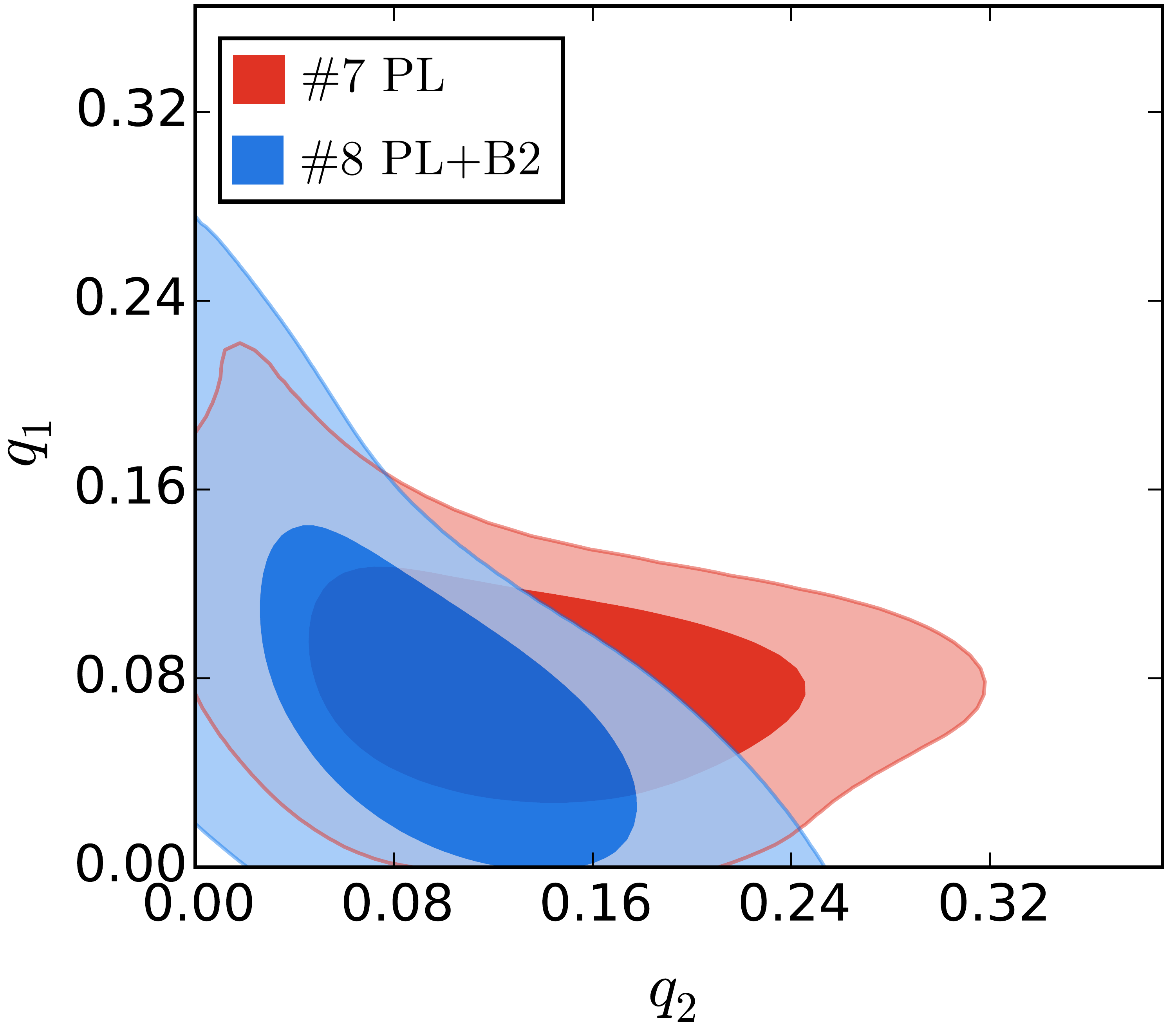, width=6.5 cm}
\caption{\it On the l.h.s. $68\%$ and $95\%$ confidence level constraints on $r$ and $n_s$ in cases $7-8$. On the r.h.s. $68\%$ and $95\%$ confidence level constraints on $q_1$ and $q_2$ in cases $7-8$ .}
\label{fig1b}
\end{figure}
Concerning the best fit to the parameters $q_1$ and $q_2$ we observe that the two datasets lead to close predictions with
\be{bestfitq}
q_1\simeq 0.73\cdot 10^{-1}\;,\;\;2q_2\simeq 0.26
\ee
when Planck data alone are considered and
\be{bestfitq2}
q_1\simeq 0.59\cdot 10^{-1}\;,\;\;2q_2\simeq 0.22
\ee
when BK data are added to the analysis. 
Correspondingly $n_s$ and $A_s$ also take very similar values for the best fit. Let us note that a larger tensor to scalar ratio w.r.t. the case 5 with a running spectral index (see figure \ref{fig1}-\ref{fig1b}) is allowed for case 7. 

The value of $q_2$ indicates a $\sim 20-25\%$ loss in power when $k$ approaches zero and the small value of $q_1$ indicates that the quantum corrections are negligible around the pivot scale $k_*$.  Let us note that the tensor to scalar ratio $r$ may be quite large at the pivot scale when the analysis is restricted to PL data (but weakly constrained). The predictions concerning the value of $r$ and the amplitude of the scalar perturbations $A_s$ fix the energy scale of inflation
\be{scaleinf}
\frac{H_*^2}{\M^2}\simeq \frac{\pi^2}{2}A_s\cdot r
\ee
and in particular, on assuming $r= 0.05$ and $\ln\pa{10^{10} A_s}=3.1$ (from the best fit of the case 8), one is led to $H^2/\M^2\sim0.05\cdot 10^{-8}$.
The extrapolation (\ref{extparam}) in the $k\rightarrow 0$ limit, on comparing with (\ref{papprox}), gives the following constraint on the physical parameters of the model:
\be{fitphys}
\frac{H^2}{\M^2}\pa{\frac{\delta_k}{k}+\frac{17}{9}-\frac{7}{18}{\rm e}^{N_*\pa{1-n_s-r/8}}}\simeq -2q_1^3 q_2\pa{\frac{k_*}{\bar k}}^3.
\ee
For reasonable values of $N_*\sim 60$, $r=0.05$, $q_1$ and $q_2$ given by (\ref{bestfitq}) and imposing the BD prescription for the vacuum, $\delta_k=-k/2$, we observe that the quantity inside the bracket on the l.h.s. of (\ref{fitphys}) is $\sim -2.5$. Let us note that slightly different values for $r$, $n_s$ and $N_*$ may lead to results which differ by an order of magnitude. 

Correspondingly, for the cases $\#7$ and $\#8$ we find the following estimate for the cut-off scale $\bar k$
\be{bestfitphys}
\bar k_{\#7}\simeq 38 k_*\simeq 1.9\;{\rm Mpc}^{-1}\, ,\;\bar k_{\#8}\simeq 30 k_*\simeq 1.6\;{\rm Mpc}^{-1}
\ee
which are very large compared to the wave number associated with the largest observable scale in the CMB namely $k_{\rm min}\simeq 1.4\cdot  10^{-4}\;{\rm Mpc}^{-1}$. Let us note that the existence of such a small fundamental scale may have relevant consequences on astrophysical observation. Indeed it is associated with distances which are comparable with the diameter of our galaxy. \\
\begin{figure}[t!]
\centering
\epsfig{file=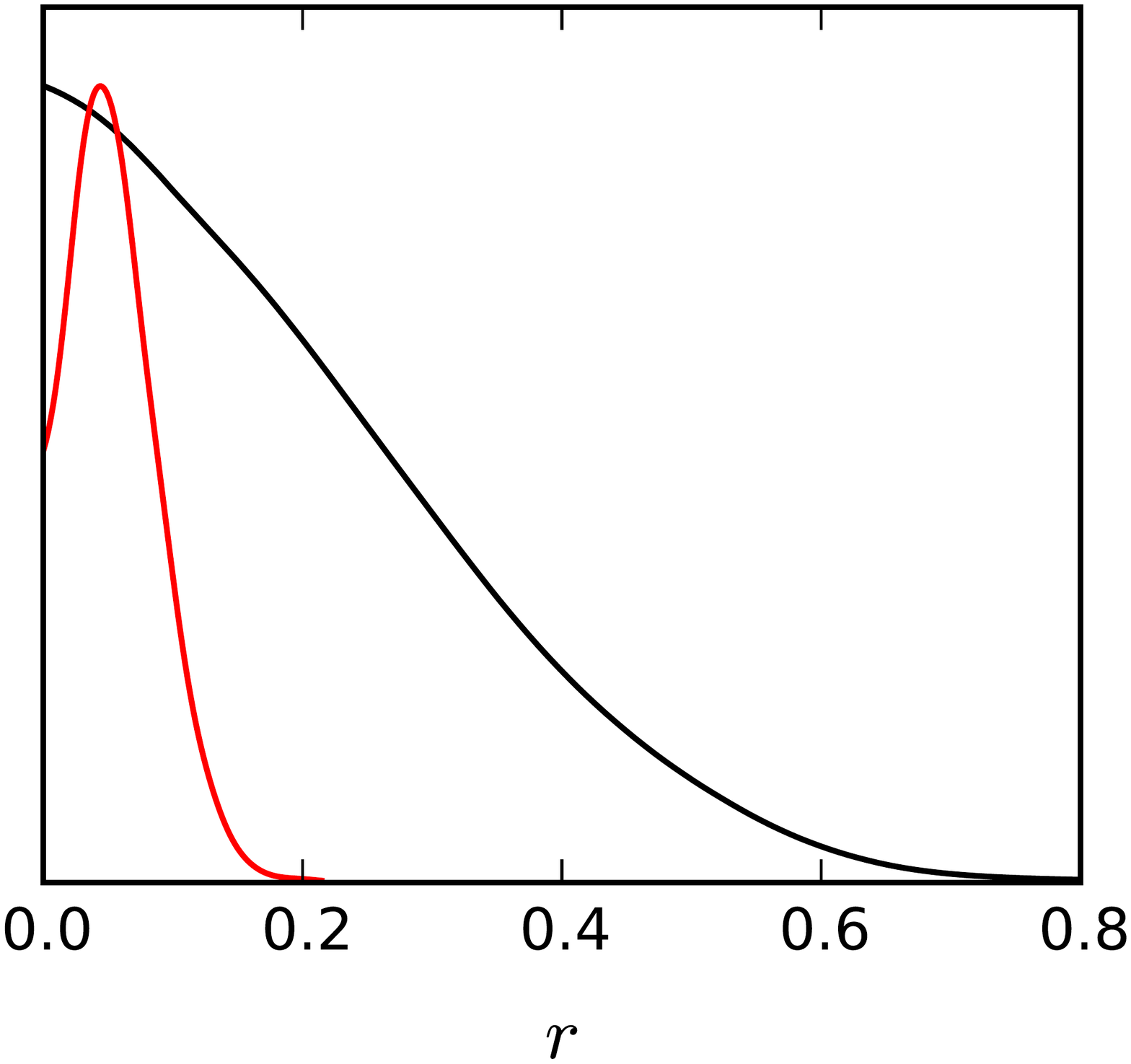, width=4.4 cm}
\epsfig{file=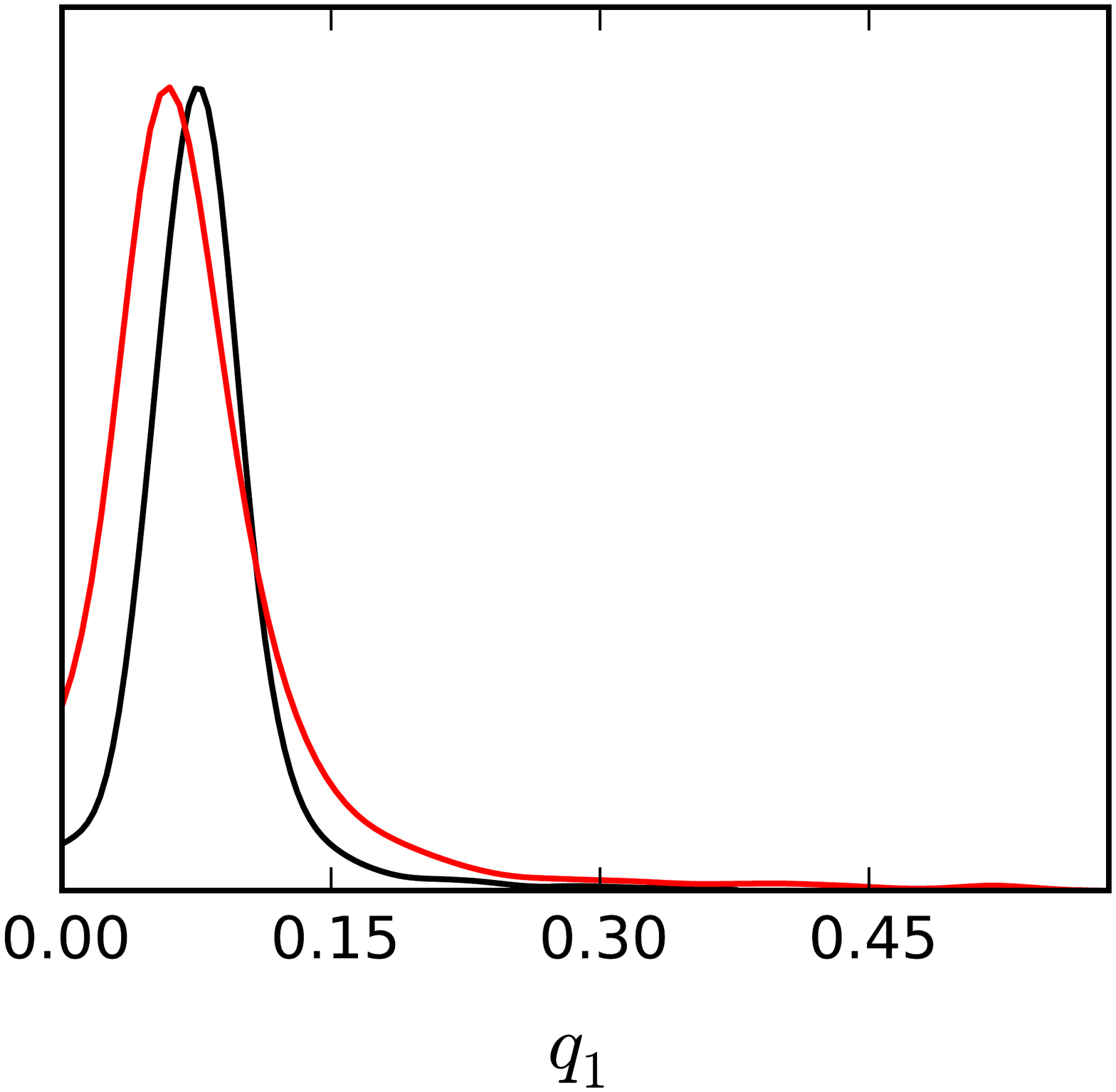, width=4.4 cm}
\epsfig{file=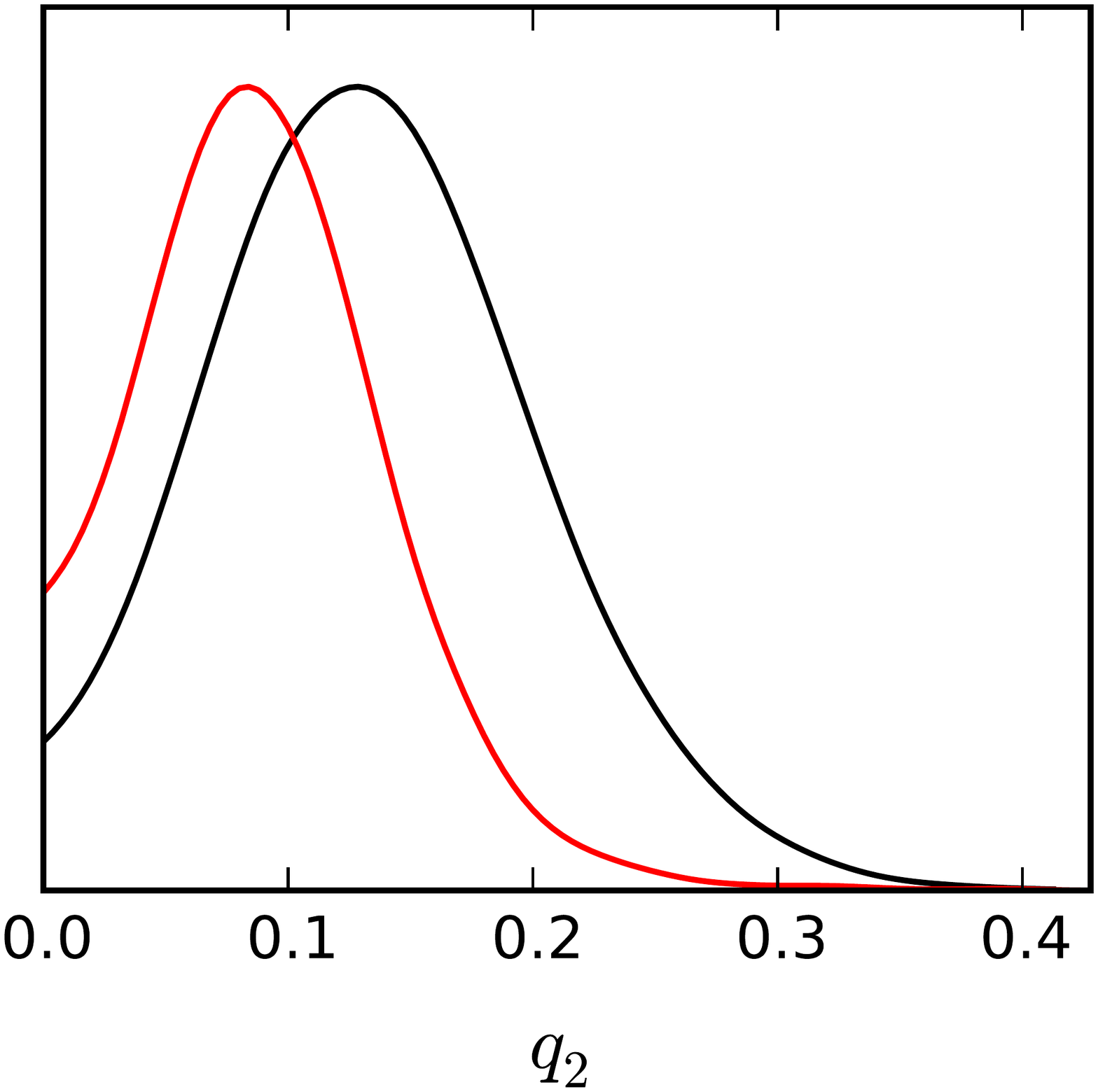, width=4.4 cm}
\caption{\it Marginalized 1-D likelihoods for $r$, $q_1$ and $q_2$ in cases $7$ (black line) and $8$ (red line).}
\label{fig3}
\end{figure}
The primordial scalar spectrum with a running spectral index (case 6) is almost indistinguishable from case $8$ in polarization (see Table \ref{tab4b} where the effective $\chi^2$ for BK data is presented) but is disfavored ($\Delta\chi^2\sim 2.6$ on adding only one free parameter to the fit) when Planck and BK data are simultaneously considered. 
In figure \ref{fig3} we finally plot the marginalized likelihoods for $r$, $q_1$ and $q_2$. The correspondent marginalized $68\%$, $95\%$ and $99\%$ confidence intervals are listed in tables \ref{tab5}-\ref{tab8} for cases 5, 6, 7 and 8 respectively.\\

\begin{table}[h!]
\caption{Marginalized confidence intervals - Case 5}
\vspace{0.1 cm}
\centering
\begin{tabular}{c | c c c}
  & $68\%$ & $95\%$& $99\%$  \\
[0.1ex]
\hline
$r$			& $[0,0.94\cdot 10^{-1}]$ 	& $[0,0.18]$ 			&  $[0,0.24]$ 	\\
$-10\cdot\alpha_s$	& $[0.78\cdot 10^{-1},0.21]$	& $[0.22\cdot 10^{-1},0.28]$ 	& $[-0.16\cdot 10^{-1},0.32]$	\\ 
[1ex]
\hline
\end{tabular}
\label{tab5}
\end{table}

\begin{table}[h!]
\caption{Marginalized confidence intervals - Case 6}
\vspace{0.1 cm}
\centering
\begin{tabular}{c | c c c}
  & $68\%$ & $95\%$& $99\%$  \\
[0.1ex]
\hline
$r$			& $[0.90\cdot 10^{-2},0.69\cdot 10^{-1}]$ 	& $[0,0.10]$ 			&  $[0,0.13]$ 	\\
$-10\cdot\alpha_s$	& $[0.67\cdot 10^{-1},0.18]$	& $[0.94 \cdot 10^{-2},0.25]$ 	& $[-0.23\cdot 10^{-1},0.28]$	\\ 
[1ex]
\hline
\end{tabular}
\label{tab6}
\end{table}

\begin{table}[h!]
\caption{Marginalized confidence intervals - Case 7}
\vspace{0.1 cm}
\centering
\begin{tabular}{c | c c c}
  & $68\%$ & $95\%$& $99\%$  \\
[0.1ex]
\hline
$r$			& $[0,0.24]$ 	& $[0,0.47]$ 			&  $[0,0.59]$ 	\\
$q_1$		& $[0.47\cdot 10^{-1},0.10]$	& $[0.66\cdot 10^{-2},0.15]$ 	& $[0,0.24]$	\\ 
$q_2$		& $[0.63\cdot 10^{-1},0.19]$	& $[0.45\cdot 10^{-2},0.26]$ 	& $[0,0.30]$	\\
[1ex]
\hline
\end{tabular}
\label{tab7}
\end{table}
\begin{table}[h!]
\caption{Marginalized confidence intervals - Case 8}
\vspace{0.1 cm}
\centering
\begin{tabular}{c | c c c}
  & $68\%$ & $95\%$& $99\%$  \\
[0.1ex]
\hline
$r$			& $[0.12\cdot 10^{-1},0.79\cdot 10^{-1}]$ 	& $[0,0.11]$ 			&  $[0,0.14]$ 	\\
$q_1$		& $[0.23\cdot 10^{-1},0.10]$	& $[0,0.19]$ 	& $[0,0.39]$	\\ 
$q_2$		& $[0.35\cdot 10^{-1},0.13]$	& $[0,0.18]$ 	& $[0,0.24]$	\\
[1ex]
\hline
\end{tabular}
\label{tab8}
\end{table}

The marginalized likelihoods for $q_1$ and $q_2$ show a $2\sigma$ deviation from standard power law for case 7 and a $1\sigma$ deviation for case 8. 
Let us note that the marginalized likelihood for $q_2$ rules out $q_2=1/2$ (i.e. the case of no $k/k_*$ dependence at the numerator in (\ref{scparam})) at more than $3\sigma$ in both case 7 and 8. Thus the comparison with the data leads to severe restrictions on the form of the quantum gravitational corrections (\ref{extparam}) as a function of $(\bar k/k)^3$. 
\section{Conclusions}
The possibility of a loss of power at large scales in the spectrum of the CMB temperature anisotropies has intriguing consequences. On the phenomenological side the coexistence of these effects signals the necessity of modifying the standard power law assumption for the primordial spectra predicted in the context of SR inflation. On the theoretical side both these effects can be related to the physics at and above the Planck scale and (even if still uncertain given the present precision of the data) should be interpreted in the context of quantum gravity.\\ 
In this paper we present an analysis of the effects of the quantum gravitational corrections on the spectra of primordial perturbations calculated in the context of a Wheeler-De Witt approach. Such effects have been obtained and discussed in detail in \cite{Kamenshchik:2014kpa} and have a quite distinctive form which affects more the large scales of the CMB. For simplicity the analysis has been restricted to the particular case of negligible quantum gravitational contributions on the spectrum of primordial gravitational waves. The NLO predictions for the scalar sector have been described in terms of two parameters, suitably extrapolated beyond the NLO and examined down to $k\rightarrow 0$. Finally, we compared PL and BK observation with our predictions by using a common parametrization for the primordial spectra and varying only a few relevant parameters in the cases discussed. Other parameterizations have been considered, however the one we present is the simplest and leads to the best results.\\
The results show that the effect of our quantum gravitational corrections compared to the standard cases leads to a better fit with the data. A similar result has been obtained in the literature \cite{powerloss} by adopting a different modification of the primordial spectra leading to a power loss at large scales and a similar improvement to the fit was found. Nonetheless the possibility of explaining the loss in power as an effect of quantum gravity on standard SR inflation is intriguing and the results obtained suggest the possibility of looking for quantum gravity effects at much smaller scales.  
On including the BK data in our analysis we found that the statistical relevance of the possibility of a loss in power for large scales is nearly invariant.
Furthermore the comparison with the data predicts, for our model, a loss in power of about $20-25\%$ w.r.t. the standard power law and fixes the scale $\bar k$ which necessarily appears in the theoretical model. If such a scale plays a crucial role in the description of gravitational phenomena one may expect that other astrophysical observables are affected around it. Unfortunately, due to the large error bars in the temperature correlations at large scales, observations are not able to severely constrain the parameter of the model. A $1\sigma-2\sigma$ deviation from the standard power law parametrization is found. Still the quite distinctive form of the quantum gravitational corrections gives a clear indication on their shape. 

\section*{Acknowledgments}
The authors thank A. Gruppuso and F. Finelli for useful comments and discussions. 
The work of A. K. was partially supported by the RFBR grant 14-02-00894.



\begin{thebibliography}{99}
\bibitem{planck}
  P.~A.~R.~Ade {\it et al.}  [Planck Collaboration],
  arXiv:1303.5082 [astro-ph.CO];
  P.~A.~R.~Ade {\it et al.}  [Planck Collaboration],
  Astron.\ Astrophys.\  {\bf 571} (2014) A15
  [arXiv:1303.5075 [astro-ph.CO]].
 \bibitem{inflation}
 A.~A.~Starobinsky,
  Phys.\ Lett.\ B {\bf 91} (1980) 99;

A.~H.~Guth,
  Phys.\ Rev.\ D {\bf 23} (1981) 347;
 A.~D.~Linde,
  Phys.\ Lett.\ B {\bf 108} (1982) 389;
  
  A.~Albrecht and P.~J.~Steinhardt,
  Phys.\ Rev.\ Lett.\  {\bf 48} (1982) 1220;
 
A. A. Starobinsky, Lect. Notes Phys. {\bf 246}, 107 (1986);
A. D. Linde, Particle Physics and Inflationary Cosmology, Harwood Chur, Switzerland, 1990. 


 \bibitem{Stewart:1993bc}
  A.~A.~Starobinsky,
  JETP Lett.\  {\bf 30} (1979) 682
   [Pisma Zh.\ Eksp.\ Teor.\ Fiz.\  {\bf 30} (1979) 719];
   
   V.~F.~Mukhanov and G.~V.~Chibisov,
  JETP Lett.\  {\bf 33} (1981) 532
   [Pisma Zh.\ Eksp.\ Teor.\ Fiz.\  {\bf 33} (1981) 549];
  
  S.~W.~Hawking and I.~G.~Moss,
  Nucl.\ Phys.\ B {\bf 224} (1983) 180;
  
  A.~A.~Starobinsky,
  Phys.\ Lett.\ B {\bf 117} (1982) 175;
  
  A.~H.~Guth and S.~Y.~Pi,
  Phys.\ Rev.\ Lett.\  {\bf 49} (1982) 1110;
   
  E.~D.~Stewart and D.~H.~Lyth,
  Phys.\ Lett.\ B {\bf 302} (1993) 171
  [gr-qc/9302019].
  
  \bibitem{WMAP}
  H.~V.~Peiris {\it et al.}  [WMAP Collaboration],
  Astrophys.\ J.\ Suppl.\  {\bf 148} (2003) 213
  [astro-ph/0302225].
 
 
 
 
 
 

\bibitem{powerloss}
C.~R.~Contaldi, M.~Peloso, L.~Kofman and A.~D.~Linde,
  JCAP {\bf 0307} (2003) 002
  [astro-ph/0303636];
  M.~Cicoli, S.~Downes, B.~Dutta, F.~G.~Pedro and A.~Westphal,
  arXiv:1407.1048 [hep-th];
  D.~K.~Hazra, A.~Shafieloo, G.~F.~Smoot and A.~A.~Starobinsky,
  JCAP {\bf 1406} (2014) 061
  [arXiv:1403.7786 [astro-ph.CO]];
 F.~G.~Pedro and A.~Westphal,
  JHEP {\bf 1404} (2014) 034
  [arXiv:1309.3413 [hep-th]];
  E.~Dudas, N.~Kitazawa, S.~P.~Patil and A.~Sagnotti,
  JCAP {\bf 1205} (2012) 012
  [arXiv:1202.6630 [hep-th]];
  M.~Cicoli, S.~Downes and B.~Dutta,
  JCAP {\bf 1312} (2013) 007
  [arXiv:1309.3412 [hep-th], arXiv:1309.3412];
  C.~R.~Contaldi, M.~Peloso and L.~Sorbo,
  JCAP {\bf 1407} (2014) 014
  [arXiv:1403.4596 [astro-ph.CO]];
  A.~Gruppuso, P.~Natoli, F.~Paci, F.~Finelli, D.~Molinari, A.~De Rosa and N.~Mandolesi,
  JCAP {\bf 1307} (2013) 047
  [arXiv:1304.5493 [astro-ph.CO]];
  A.~Gruppuso,
  Mon.\ Not.\ Roy.\ Astron.\ Soc.\  {\bf 437} (2014) 2076
  [arXiv:1310.2822 [astro-ph.CO]];
  
  
  
  
\bibitem{quantumloss}
A.~Y.~Kamenshchik, A.~Tronconi and G.~Venturi,
  Phys.\ Lett.\ B {\bf 726} (2013) 518;
  G.~Calcagni,
  Annalen Phys.\  {\bf 525} (2013) 323
  [arXiv:1209.0473 [gr-qc]];
C. Kiefer and M. Kr\"amer, Phys. Rev. Lett. {\bf 108}, 021301 (2012);
D.~Bini, G.~Esposito, C.~Kiefer, M.~Kraemer and F.~Pessina,
  Phys.\ Rev.\ D {\bf 87} (2013) 104008
  [arXiv:1303.0531 [gr-qc]];
G.~L.~Alberghi, R.~Casadio, A.~Tronconi,
  Phys.\ Rev.\ D {\bf 74} (2006) 103501;

\bibitem{Kamenshchik:2014kpa}
  A.~Y.~Kamenshchik, A.~Tronconi and G.~Venturi,
  Phys.\ Lett.\ B {\bf 734} (2014) 72
  [arXiv:1403.2961 [gr-qc]].


\bibitem{BO}
M. Born and J.R. Oppenheimer, Ann. Physik {\bf 84}, 457 (1927); 
C. A. Mead and D. G. Truhlar, J. Chem. Phys. {\bf 70}, 2284 (1979);
C. A. Mead, Chem. Phys {\bf 49}, 23 (1980)
C. A. Mead, Chem. Phys {\bf 49}, 33 (1980)
R. Brout and G. Venturi, Phys. Rev. D {\bf 39}, 2436 (1989);
C.~Bertoni, F.~Finelli and G.~Venturi,
   Class.\ Quant.\ Grav.\  {\bf 13}, 2375 (1996);
G. Venturi, Class. Quantum Grav. {\bf 7}, 1075 (1990);
G.~L.~Alberghi, C.~Appignani, R.~Casadio, F.~Sbisa, A.~Tronconi,
  Phys.\ Rev.\ D {\bf 77} (2008) 044002;
 A.~Tronconi, G.~P.~Vacca and G.~Venturi,
  Phys.\ Rev.\ D {\bf 67}, 063517 (2003).


\bibitem{MukMald}
V.F. Mukhanov, Sov. Phys. JETP {\bf 68}, 1297 (1988);
J.~M.~Maldacena,
  JHEP {\bf 0305} (2003) 013;
 V.~F.~Mukhanov, H.~A.~Feldman and R.~H.~Brandenberger,
  Phys.\ Rept.\  {\bf 215} (1992) 203.
 V.F. Mukhanov, Phys. Lett. B {\bf 218}, 17 (1989). 

\bibitem{DeWitt}
B.S. DeWitt, Phys. Rev. {\bf 160}, 113 (1967).

  \bibitem{BD}
  T.~S.~Bunch and P.~C.~W.~Davies,
  Proc.\ Roy.\ Soc.\ Lond.\ A {\bf 360} (1978) 117. 

\bibitem{FVV}
F. Finelli, G.P. Vacca and G. Venturi, Phys. Rev. D {\bf 58}, 103514 (1998).

\bibitem{cosmomc}
  A.~Lewis and S.~Bridle,
  Phys.\ Rev.\ D {\bf 66} (2002) 103511
  [astro-ph/0205436].

 \bibitem{bicep}
  P.~A.~R.~Ade {\it et al.}  [BICEP2 and Planck Collaborations],
  Phys.\ Rev.\ Lett.\  {\bf 114} (2015) 10,  101301
  [arXiv:1502.00612 [astro-ph.CO]].
  
 \bibitem{hert}
T.~Hertog,
  JCAP {\bf 1402} (2014) 043
  [arXiv:1305.6135 [astro-ph.CO]].





\end{thebibliography}
\end{document}